\pdfoutput=1
\documentclass[12pt]{article}

\usepackage{graphicx}

\input{epsf}

\advance \topmargin by -\headheight
\advance \topmargin by -\headsep

\evensidemargin \oddsidemargin
\begin{document}

\topmargin -.6in

\def\rh{{\hat \rho}}
\def\alie{{\hat{\cal G}}}
\newcommand{\sect}[1]{\setcounter{equation}{0}\section{#1}}
\renewcommand{\theequation}{\thesection.\arabic{equation}}

\def\rf#1{(\ref{eq:#1})}
\def\lab#1{\label{eq:#1}}
\def\nonu{\nonumber}
\def\br{\begin{eqnarray}}
\def\er{\end{eqnarray}}
\def\be{\begin{equation}}
\def\ee{\end{equation}}
\def\eq{\!\!\!\! &=& \!\!\!\! }
\def\foot#1{\footnotemark\footnotetext{#1}}
\def\lb{\lbrack}
\def\rb{\rbrack}
\def\llangle{\left\langle}
\def\rrangle{\right\rangle}
\def\blangle{\Bigl\langle}
\def\brangle{\Bigr\rangle}
\def\llbrack{\left\lbrack}
\def\rrbrack{\right\rbrack}
\def\lcurl{\left\{}
\def\rcurl{\right\}}
\def\({\left(}
\def\){\right)}
\newcommand{\nit}{\noindent}
\newcommand{\ct}[1]{\cite{#1}}
\newcommand{\bi}[1]{\bibitem{#1}}
\def\lskip{\vskip\baselineskip\vskip-\parskip\noindent}
\relax

\def\tr{\mathop{\rm tr}}
\def\Tr{\mathop{\rm Tr}}
\def\trace{\widehat{\rm Tr}}
\def\v{\vert}
\def\bv{\bigm\vert}
\def\Bgv{\;\Bigg\vert}
\def\bgv{\bigg\vert}
\newcommand\partder[2]{{{\partial {#1}}\over{\partial {#2}}}}
\newcommand\funcder[2]{{{\delta {#1}}\over{\delta {#2}}}}
\newcommand\Bil[2]{\Bigl\langle {#1} \Bigg\vert {#2} \Bigr\rangle}  
\newcommand\bil[2]{\left\langle {#1} \bigg\vert {#2} \right\rangle} 
\newcommand\me[2]{\left\langle {#1}\bv {#2} \right\rangle} 
\newcommand\sbr[2]{\left\lbrack\,{#1}\, ,\,{#2}\,\right\rbrack}
\newcommand\pbr[2]{\{\,{#1}\, ,\,{#2}\,\}}
\newcommand\pbbr[2]{\lcurl\,{#1}\, ,\,{#2}\,\rcurl}

\def\ket#1{\mid {#1} \rangle}
\def\bra#1{\langle {#1} \mid}
\newcommand{\braket}[2]{\langle {#1} \mid {#2}\rangle}
%
\def\a{\alpha}
\def\at{{\tilde A}^R}
\def\atc{{\tilde {\cal A}}^R}
\def\atcm#1{{\tilde {\cal A}}^{(R,#1)}}
\def\b{\beta}
\def\dc{{\cal D}}
\def\d{\delta}
\def\D{\Delta}
\def\eps{\epsilon}
\def\vareps{\varepsilon}
\def\g{\gamma}
\def\G{\Gamma}
\def\grad{\nabla}
\def\h{{1\over 2}}
\def\l{\lambda}
\def\L{\Lambda}
\def\m{\mu}
\def\n{\nu}
\def\o{\over}
\def\om{\omega}
\def\O{\Omega}
\def\p{\phi}
\def\P{\Phi}
\def\pa{\partial}
\def\pr{\prime}
\def\pt{{\tilde \Phi}}
\def\qs{Q_{\bf s}}
\def\ra{\rightarrow}
\def\s{\sigma}
\def\S{\Sigma}
\def\t{\tau}
\def\th{\theta}
\def\Th{\Theta}
\def\tpp{\Theta_{+}}
\def\tmm{\Theta_{-}}
\def\tpg{\Theta_{+}^{>}}
\def\tms{\Theta_{-}^{<}}
\def\tp0{\Theta_{+}^{(0)}}
\def\tm0{\Theta_{-}^{(0)}}
\def\ti{\tilde}
\def\wti{\widetilde}
\def\jc{J^C}
\def\bj{{\bar J}}
\def\sj{{\jmath}}
\def\bsj{{\bar \jmath}}
\def\bp{{\bar \p}}
\def\vp{\varphi}
\def\ve{\varepsilon}
\def\vt{{\tilde \varphi}}
\def\faa{Fa\'a di Bruno~}
\def\ca{{\cal A}}
\def\cb{{\cal B}}
\def\ce{{\cal E}}
\def\cg{{\cal G}}
\def\cgh{{\hat {\cal G}}}
\def\ch{{\cal H}}
\def\chh{{\hat {\cal H}}}
\def\cl{{\cal L}}
\def\cm{{\cal M}}
\def\cn{{\cal N}}
\def\u2{\mid u\mid^2}
\def\ub{{\bar u}}
\def\z2{\mid z\mid^2}
\def\zb{{\bar z}}
\def\w2{\mid w\mid^2}
\def\wb{{\bar w}}
\newcommand\sumi[1]{\sum_{#1}^{\infty}}   
\newcommand\fourmat[4]{\left(\begin{array}{cc}  
{#1} & {#2} \\ {#3} & {#4} \end{array} \right)}

%
\def\lie{{\cal G}}
\def\kmlie{{\hat{\cal G}}}
\def\dlie{{\cal G}^{\ast}}
\def\elie{{\widetilde \lie}}
\def\edlie{{\elie}^{\ast}}
\def\hlie{{\cal H}}
\def\flie{{\cal F}}
\def\wlie{{\widetilde \lie}}
\def\f#1#2#3 {f^{#1#2}_{#3}}
\def\winf{{\sf w_\infty}}
\def\win1{{\sf w_{1+\infty}}}
\def\hwinf{{\sf {\hat w}_{\infty}}}
\def\Winf{{\sf W_\infty}}
\def\Win1{{\sf W_{1+\infty}}}
\def\hWinf{{\sf {\hat W}_{\infty}}}
\def\Rm#1#2{r(\vec{#1},\vec{#2})}          
\def\OR#1{{\cal O}(R_{#1})}           
\def\ORti{{\cal O}({\widetilde R})}           
\def\AdR#1{Ad_{R_{#1}}}              
\def\dAdR#1{Ad_{R_{#1}^{\ast}}}      
\def\adR#1{ad_{R_{#1}^{\ast}}}       
\def\KP{${\rm \, KP\,}$}                 
\def\KPl{${\rm \,KP}_{\ell}\,$}         
\def\KPo{${\rm \,KP}_{\ell = 0}\,$}         
\def\mKPa{${\rm \,KP}_{\ell = 1}\,$}    
\def\mKPb{${\rm \,KP}_{\ell = 2}\,$}    
%
\def\rlx{\relax\leavevmode}
\def\inbar{\vrule height1.5ex width.4pt depth0pt}
\def\IZ{\rlx\hbox{\sf Z\kern-.4em Z}}
\def\IR{\rlx\hbox{\rm I\kern-.18em R}}
\def\IC{\rlx\hbox{\,$\inbar\kern-.3em{\rm C}$}}
\def\IN{\rlx\hbox{\rm I\kern-.18em N}}
\def\IO{\rlx\hbox{\,$\inbar\kern-.3em{\rm O}$}}
\def\IP{\rlx\hbox{\rm I\kern-.18em P}}
\def\IQ{\rlx\hbox{\,$\inbar\kern-.3em{\rm Q}$}}
\def\IF{\rlx\hbox{\rm I\kern-.18em F}}
\def\IG{\rlx\hbox{\,$\inbar\kern-.3em{\rm G}$}}
\def\IH{\rlx\hbox{\rm I\kern-.18em H}}
\def\II{\rlx\hbox{\rm I\kern-.18em I}}
\def\IK{\rlx\hbox{\rm I\kern-.18em K}}
\def\IL{\rlx\hbox{\rm I\kern-.18em L}}
\def\one{\hbox{{1}\kern-.25em\hbox{l}}}
\def\0#1{\relax\ifmmode\mathaccent"7017{#1}%
B        \else\accent23#1\relax\fi}
\def\omz{\0 \omega}
%
\def\ltimes{\mathrel{\vrule height1ex}\joinrel\mathrel\times}
\def\rtimes{\mathrel\times\joinrel\mathrel{\vrule height1ex}}
%
\def\mark{\noindent{\bf Remark.}\quad}
\def\prop{\noindent{\bf Proposition.}\quad}
\def\theor{\noindent{\bf Theorem.}\quad}
\def\name{\noindent{\bf Definition.}\quad}
\def\exam{\noindent{\bf Example.}\quad}
\def\proof{\noindent{\bf Proof.}\quad}

\begin{titlepage}
\vspace*{-1cm}

\vskip 3cm

\vspace{.2in}
\begin{center}
{\large\bf  Self-Duality in the Context of the  Skyrme Model }
\end{center}

\vspace{.5cm}

\begin{center}
L. A. Ferreira\footnote{laf@ifsc.usp.br} and L. R. Livramento\footnote{leandrorl@ifsc.usp.br}

\vspace{.3 in}
\small

\par \vskip .2in \noindent
Instituto de F\'\i sica de S\~ao Carlos; IFSC/USP;\\
Universidade de S\~ao Paulo, USP  \\ 
Caixa Postal 369, CEP 13560-970, S\~ao Carlos-SP, Brazil\\

\normalsize
\end{center}

\vspace{.5in}

\begin{abstract}

 We study a recently proposed modification of the Skyrme model that possesses an exact self-dual sector leading to an infinity of exact Skyrmion solutions with arbitrary topological (baryon) charge.  The self-dual sector is made possible by the introduction, in addition to the usual  three $SU(2)$ Skyrme fields, of six scalar fields assembled in a symmetric and invertible three dimensional matrix $h$. The action presents   quadratic and quartic terms in derivatives of the Skyrme fields, but instead of the group indices being contracted by the $SU(2)$  Killing form, they are contracted with the $h$-matrix in the quadratic term, and by its inverse on the quartic term. Due to these extra fields the static version of the model, as well as its self-duality equations, are conformally invariant on the three dimensional space $\IR^3$.  We show that the static and self-dual sectors of such a theory are equivalent, and so the only non-self-dual solution must be time dependent. We also show that for any configuration of the Skyrme $SU(2)$ fields, the $h$-fields adjust themselves to satisfy the self-duality equations, and so the theory has plenty of non-trivial topological solutions. We present explicit exact solutions using a holomorphic rational ansatz, as  well as a toroidal ansatz based on the conformal symmetry. We point to possible extensions of the model that break the conformal symmetry as well as the self-dual sector, and that can perhaps lead to interesting physical applications.

\end{abstract} 
\end{titlepage}

\section{Introduction}
\label{sec:intro}
\setcounter{equation}{0}

The concept of self-duality plays a crucial role in the study of classical solutions of a great variety of field theories, from  kinks in $(1+1)$-dimensions to instantons and magnetic monopoles in four dimensional non-abelian gauge theories. The appearance of self-dual solutions depends crucially on the existence in the theory, of a topological charge admitting an integral representation, i.e. there must exists a local density of topological charge. The self-dual solutions present two main properties: first, they are solutions of self-duality equations which are first order (partial) differential equations that imply the (static) second order Euler-Lagrange equations of the full theory, and second, they saturate a lower bound of the static energy (or Euclidean action) which is related to the modulus of the topological charge. This second property makes the self-dual solutions very stable as they have the lowest  energy in any topological sector of the theory. The fact that the construction of self-dual solutions involve one integration less than the case of the ordinary solutions of the Euler-Lagrange equations, is not related to the use of dynamical conservation laws, but a consequence of the topological structures of the theory. Indeed, the invariance of the topological charge under any smooth (homotopic) variation of the fields leads,  with its integral representation,  to local differential identities that together with the self-duality equations imply the Euler-Lagrange equations \cite{selfdual}. 

In this paper we explore the concept of self-duality in the context of the Skyrme model \cite{skyrme}, a $(3+1)$-dimensional field theory with fields taking values on the group $SU(2)$ and presenting a lot of interesting physical applications \cite{mantonbook,shnirbook}. Despite the fact that the Skyrme model does have a topological charge admitting an integral representation, it does not possess a non-trivial self-dual sector \cite{ruback}. The study of the properties of its topological solitons, the Skyrmions, has therefore to rely on numerical methods. Even though that has not prevented a large number of physical applications, the existence of a self-dual sector could shed light on many of the structures of the model.  In the last few years there has appeared some modifications of the Skyrme model that do admit  self-dual sectors. There is the approach, based of self-dual Yang-Mills, that amounts to the coupling an infinite tower of meson fields to the Skyrme model \cite{sutcliffetower} and that has obtained very interesting results for the spectrum of light nuclei \cite{naya1,naya2}. There is the so-called BPS Skyrme model \cite{adam1,adam2} that instead of the usual quadratic and quartic Skyrme terms in derivatives of the fields, it has in its action a term with six derivatives of the fields (the square of the topological current) and a potential term. The Skyrmions in such a model are constructed analytically and are compacton-like solutions.  Some applications in nuclear physics and neutron stars were achieved through such a model \cite{adamprl,adamneutron}. Then there is a modification of the Skyrme model \cite{ua845}, using the ideas of \cite{selfdual}, that  possesses an exact self-dual sector, and when coupled to an extra field presents an infinite number of analytical self-dual solutions   \cite{shnir} with arbitrary values of the Skyrme topological charge. 

The model we consider in this paper is the one proposed in \cite{laf}, possessing an exact self-dual sector, and also constructed based on the ideas of \cite{selfdual}. Besides the usual $SU(2)$ group valued fields $U$ of the Skyrme model, it has six scalar fields assembled in a symmetric and invertible matrix $h_{ab}$, $a,b=1,2,3$. The action is similar to that of the original Skyrme model, in the sense it has  quadratic and quartic terms in derivatives of the $U$-fields, but the group indices are not contracted with the $SU(2)$ Killing (trace) form, but with the matrix  $h_{ab}$ in the quadratic term, and its inverse in the quartic term. The model is defined by the action
\be
S= \int d^4x\left[ \frac{m_0^2}{2}\, h_{ab}\,R^a_{\mu}\,R^{b\,,\, \mu}-\frac{1}{4\,e_0^2}\, h^{-1}_{ab}\,H^a_{\mu\nu}\,H^{b\,,\,\mu\nu}\right]
\lab{modelintro}
\ee
where, like in the usual Skyrme model, $R^a_{\mu}$ are the components of the Maurer-Cartan form, i.e. $i\,\partial_{\mu}U\,U^{\dagger}\equiv R^a_{\mu}\,T_a$, with $T_a$ being a basis of the $SU(2)$ Lie algebra, and $H^a_{\mu\nu}$ is the curl of that form, i.e. $H^a_{\mu\nu}\equiv \partial_{\mu} R^a_{\nu}-\partial_{\nu} R^a_{\mu}$, and $m_0$ and $e_0$ are coupling constants. Of course, in order to keep the energy of the theory \rf{modelintro} positive definite, we shall restrict the matrix $h_{ab}$ to the cases where its eigenvalues   are positive. If the fields $h_{ab}$ are considered as independent fields, then their presence does not destroy, as we discuss in section \ref{sec:model}, the usual global left and right symmetry $SU(2)_L\otimes SU(2)_R$ of the original Skyrme model \cite{skyrme}. The introduction of the extra fields $h_{ab}$ is motivated by the methods of \cite{selfdual} of constructing theories with self-dual sectors. The method involves the splitting of the density of the topological charge into two pieces and the static energy is built by adding the squares of these quantities. In the splitting process  one is free to attach a matrix to one of the pieces and its inverse to the other piece. The self-duality equations is obtained by imposing the equality, up to a sign, of these two pieces. In the present case the topological charge used in such a  process is the usual Skyrme topological charge associated   to the mappings $\IR^3 \rightarrow SU(2)$, with the spatial infinity identified to a point, and so with $\IR^3$ compactified to the $3$-sphere $S^3$. The corresponding self-duality equations are given by
\be
m_0\,e_0 \,h_{ab}\,R^b_i=\pm \frac{1}{2}\,\ve_{ijk}\,H^a_{jk}
\lab{selfdualeqsintro}
\ee
with the indices $i,j,k=1,2,3$, corresponding to the spatial Cartesian coordinates $x^i$. Any solution of the nine first order partial differential equations \rf{selfdualeqsintro} also solves not only the static (second order) Euler-Lagrange equations associated to the $SU(2)$ $U$-fields, but also the static Euler-Lagrange equations associated to the extra fields $h_{ab}$. Another interesting point about the presence of the fields $h_{ab}$ is that they render the static energy, associated to the theory \rf{modelintro}, and the self-duality equations \rf{selfdualeqsintro}, invariant under conformal transformations in the three dimensional spatial sub-manifold $\IR^3$. As a consequence, the self-dual Skyrmions do not have a fixed size, but can be re-scaled without changing its energy and topological charge. However, the equality of the contributions to the static energy coming from the quadratic and quartic terms, which in the original Skyrme model is implied by Derrick's scaling argument  \cite{derrick}, here is a consequence of the static Euler-Lagrange equations associated to the fields $h_{ab}$, as we discuss in section \ref{sec:model}. 

In \cite{laf} it was constructed analytical self-dual Skyrmions of unity topological charge for the theory \rf{modelintro}, in the case where the matrix $h_{ab}$ is proportional to the unity matrix, and so one has just one extra field. In this paper we analyse the properties of the theory \rf{modelintro} in its full generality. The first two important results that we obtain are the following. The static sector of the theory \rf{modelintro} is the same as its self-dual sector, i.e. any static solution of \rf{modelintro} is a solution of the self-duality equations \rf{selfdualeqsintro}. The static Euler-Lagrangian equations associated to the fields $h_{ab}$ play a crucial role in establishing such a result, and in fact they are in some sense equivalent to the self-duality equations \rf{selfdualeqsintro}. In addition, the fields $h_{ab}$ are just spectators in the self-dual or static sector, in the sense that given any configuration for the $SU(2)$ $U$-fields, the fields $h_{ab}$ adjust themselves to satisfy all the nine self-duality equations \rf{selfdualeqsintro}. In fact, for any self-dual solution,    the fields $h_{ab}$ can be written as 
\be
h=\frac{\sqrt{{\rm det}\,\tau}}{\mid m_0\,e_0\mid}\; \tau^{-1}
\lab{goodrelhtauintro}
\ee
where $\tau$ depends only on the $U$-fields and it is given by  $\tau_{ab}= R_{i}^a\,R_{i}^b$.  Note that $\tau$ is similar to the strain tensor of the Skyrme model given by $D_{ij}=R_i^a\,R_j^a$, and indeed we show that their eigenvalues are the same. Therefore, the self-duality equations \rf{selfdualeqsintro} has plenty of analytical solutions, and we analyse two types of such solutions. We construct explicitly self-dual solutions for any value of the topological charge using the holomorphic rational ansatz for the $SU(2)$ $U$-fields \cite{mantonbook}. For those  configurations we construct the matrix $h_{ab}$, in particular its eigenvalues $\vp_a$. It turns out that the first two eigenvalues $\vp_1$ and $\vp_2$, are equal and spherically symmetric. They have its maximum value at the origin and decay to zero at infinity. The third eigenvalue $\vp_3$ decomposes into the product of two pieces, one  depending on the radial distance only, and the other on the  polar and azimuthal angles. The radial part resembles very much the form of $\vp_1$ and $\vp_2$, and the angular part is proportional to the Wronskian of the two holomorphic polynomials entering in the rational map. As usual the topological charge of the solution is determined by the degrees of these polynomials.  

We then use the conformal symmetry of the static version of the theory \rf{modelintro}, to construct an ansatz involving the toroidal coordinates in $\IR^3$. The symmetries of the ansatz are such that the dependence upon the two toroidal angles are explicitly given and the matrix $h_{ab}$ is given in terms of a profile function of the third toroidal coordinate, which is  left undetermined. Again the eigenvalues of $h$ have a maximum at the origin and decay to zero at spatial infinity. The topological charge is equal to the product of two integers associated in the ansatz, with the two toroidal angles. We have an infinity of analytical solutions with arbitrary topological charges. 

The construction of the model \rf{modelintro}, based on the  ideas of \cite{selfdual}, is such that the scalar fields $h_{ab}$ are introduced without a kinetic term, i.e. they are not propagating fields. In addition, it makes the static version of the theory \rf{modelintro} conformally invariant which is not very suitable for physical applications. Therefore, in order to make the model \rf{modelintro} more realistic we have to introduce a kinetic term for the fields $h_{ab}$ and break the conformal symmetry. The addition of the kinetic term alone is sufficient to break that symmetry explicitly, but one can also add  mass and  potential terms for those fields as well. The self-dual sector is also lost with the addition of such terms, but that might be  desirable for some physical applications. For instance, in applications to nuclear physics, the topological charge of the Skyrme model is interpreted as the baryon number and the Skyrmions as nuclei. In a self-dual theory the static energy of the Skyrmion is proportional  to the topological charge, and so there is no binding energy.   We have checked that the addition of the kinetic, mass and potential terms to the model \rf{modelintro} brings a positive binding energy, at least for small barion numbers. Therefore,  such modifications of the theory \rf{modelintro} seems to be promissing in applications to nuclear physics and we shall report elsewhere \cite{tocome} the numerical results we have obtained on those lines. 

The paper is organised as follows. In section \ref{sec:model} we discuss in details the properties of the model \rf{modelintro}, its symmetries, the static and self-dual sectors. In section \ref{sec:ansatzholom} we present the self-dual solutions using the rational map, and in section \ref{sec:ansatztoro} the self-dual solution in a toroidal ansatz based on the conformal symmetry. The appendix \ref{app:conformal} contains the proof of the conformal invariance of the static version of the theory \rf{modelintro}, and in appendix \ref{app:proof} we prove some results important in  establishing the equivalence of the static and self-dual sectors. The conclusions are presented in section  \ref{sec:conclusion}.

\section{The description of the model}
\label{sec:model}
\setcounter{equation}{0}

We consider in this paper the Skyrme-type model proposed in \cite{laf}, on a four dimensional Minkowski space-time, with the metric signature as $ds^2=dx_0^2-dx_i^2$, $i=1,2,3$, defined by the action 
\be
S= \int d^4x\left[ \frac{m_0^2}{2}\, h_{ab}\,R^a_{\mu}\,R^{b\,,\, \mu}-\frac{1}{4\,e_0^2}\, h^{-1}_{ab}\,H^a_{\mu\nu}\,H^{b\,,\,\mu\nu}\right]
\lab{model}
\ee
where $m_0$ (of dimension $\({\rm length}\)^{-1}$) and $e_0$ (dimensionless) are coupling constants, and  it is based on the $SU(2)$ Lie algebra with  
generators $T_a$, $a=1,2,3$,  satisfying 
\be
\sbr{T_a}{T_b}=i\,\ve_{abc}\,T_c \; ;\; \qquad\qquad\qquad\qquad 
{\rm Tr}\(T_a\,T_b\)=\kappa\, \delta_{ab}
\lab{su2killing}
\ee
where $\kappa$ is a constant that depends upon the representation ($\kappa=1/2$ for the spinor representation, and $\kappa=2$ for the triplet (adjoint) representation). We shall use a normalized trace, independent of the representation, as follows
\be
\trace\(T_a\,T_b\)=\frac{1}{\kappa}\,{\rm Tr}\(T_a\,T_b\)=\delta_{ab}
\lab{normtrace}
\ee
In addition, 
$R^a_{\mu}$, $\mu=0,1,2,3$, are the components of the $SU(2)$ Maurer-Cartan form given by
\be
R_{\mu} \equiv i\,\partial_{\mu}U\,U^{\dagger}\equiv R^a_{\mu}\,T_a\;; \qquad\qquad\qquad 
R^a_{\mu}=i\, \trace\(\partial_{\mu}U\,U^{\dagger}\,T_a\)
\lab{rdef}
\ee
with $U$ being an element of the group $SU(2)$. The quantities $H_{\mu\nu}^a$ correspond to the curl of  $R_{\mu}$, and since these satisfy the Maurer-Cartan equation,  i.e.
\be
\partial_{\mu}R_{\nu}-\partial_{\nu}R_{\mu}+i\,\sbr{R_{\mu}}{R_{\nu}}=0
\lab{maurercartaneq}
\ee
we  have that
\be
H^a_{\mu\nu}\equiv \partial_{\mu} R^a_{\nu}-\partial_{\nu} R^a_{\mu}
=-i\, \trace\(\sbr{R_{\mu}}{R_{\nu}}\,T_a\)=\ve_{abc}\,R_{\mu}^b\,R_{\nu}^c
\lab{hdef}
\ee
The theory \rf{model} differs from the original Skyrme model  \cite{skyrme,mantonbook} by the fact that the group indices are not contracted by the $SU(2)$ Killing form but instead by the symmetric matrix $h_{ab}$ on the quadratic term in derivatives and by its inverse on the quartic term. We shall consider the six entries of  that symmetric and invertible matrix  as extra fields of the theory, that  transform as scalars fields under the Lorentz  group. Clearly, we shall be concerned with  the cases where the eigenvalues of the matrix $h_{ab}$ are positive definite in order for the energy of the theory \rf{model} to be positive definite. 

Since the theory \rf{model} does not have local gauge symmetries,  in order to have finite energy solutions the fields have to go to a constant vacuum configuration at spatial infinity. Then as long as topological considerations are concerned we can compactify the space $\IR^3$ into the three-sphere $S^3$, and the fields $U$ define mappings $S^3 \rightarrow SU(2)\equiv S^3$. The corresponding topological charge (winding number)  is the same as in the usual Skyrme theory, i.e.  
\be
 Q= \frac{i}{48\,\pi^2}\int d^3x\; \ve_{ijk}\,\trace\(R_i\,R_j\,R_k\)
 \lab{topcharge}
 \ee

The Euler-Lagrange equations, obtained from \rf{model},  associated to the $SU(2)$ $U$-fields,  are given by
\be
\partial_{\mu}\left[m_0^2\,e_0^2\,h_{ab}\,R^{b,\mu}-\ve_{abc}\,R^b_{\nu}\,h_{cd}^{-1}\,H^{d,\mu\nu}\right]=\ve_{abc}\left[m_0^2\,e_0^2\,R_{\mu}^b\,h_{cd}\,R^{d,\mu}+\frac{1}{2}\,H_{\mu\nu}^c\,h_{bd}^{-1}\,H^{d,\mu\nu}\right]
\lab{elforu}
\ee
and the Euler-Lagrange equations associated to the fields $h_{ab}$ are given by
\be
m_0^2\,e_0^2\,R_{\mu}^a\,R^{b,\mu}+\frac{1}{2}\,h_{ac}^{-1}\,H_{\mu\nu}^c\,h_{bd}^{-1}\,H^{d,\mu\nu}=0
\lab{elforh}
\ee
Note that by contracting \rf{elforh} with $h_{ca}$ one gets  that the r.h.s. of \rf{elforu} must vanish. Therefore, one gets three conserved currents ($a=1,2,3$)
\be
\partial_{\mu}J^{\mu}_a=0\;;\qquad \qquad {\rm with} \qquad \qquad 
J^{\mu}_a=m_0^2\,e_0^2\,h_{ab}\,R^{b,\mu}-\ve_{abc}\,R^b_{\nu}\,h_{cd}^{-1}\,H^{d,\mu\nu}
\lab{conscurr}
\ee

The action \rf{model} is invariant under the global symmetry $SU(2)_L\otimes SU(2)_R$ defined by the transformations 
\be
U\rightarrow g_L\,U\;;\qquad\qquad R_{\mu}^a\rightarrow d_{ab}\(g_L\)\,R_{\mu}^b \;;\qquad\qquad 
h_{ab}\rightarrow d_{ac}\(g_L\)\, h_{cd}\,d_{db}^T\(g_L\)
\lab{leftsymm}
\ee
and
\be
U\rightarrow U\,g_R\;;\qquad\qquad\qquad R_{\mu}^a\rightarrow R_{\mu}^a \;;\qquad\qquad \qquad
h_{ab}\rightarrow h_{ab}
\lab{rightsymm} 
\ee
with $g_L\,,\,g_R\in SU(2)$, and where $d\(g\)$ is the $3\times 3$ matrix for the group element $g$ in the adjoint representation of $SU(2)$, i.e. 
\be
g\,T_a\, g^{-1}=T_b\,d_{ba}\(g\)\;;\qquad\qquad\qquad d\(g_1\)\,d\(g_2\)=d\(g_1\,g_2\)
\lab{adjoint}
\ee
The Noether currents associated to the left symmetry \rf{leftsymm} are those given in \rf{conscurr}, and the Noether currents associated to the right symmetry \rf{rightsymm} are given by 
\be
{\tilde J}_{\mu}^a= J_{\mu}^b\,d_{ba}\(U\) ;\qquad\qquad\qquad {\rm with}\qquad\qquad 
\partial^{\mu}{\tilde J}_{\mu}^a=0
\lab{rightcurr}
\ee
The conservation of the left and right currents, given in \rf{conscurr} and \rf{rightcurr} respectively, imply that
\be
J^{\mu}_b\,\partial_{\mu}d_{ba}\(U\)=0
\lab{funnyrel}
\ee
The relation \rf{funnyrel} is a consequence of the Euler-Lagrange equations \rf{elforh} for the fields $h_{ab}$. Indeed, using the fact that the adjoint representation is unitary and real and so the matrices $d_{ab}$ are orthogonal, i.e. $d^T\(U\)=d\(U^{\dagger}\)$, one can show that
\br
J^{\mu}_b\,\partial_{\mu}d_{ba}\(U\)\,T_a&=& 
-i\, U^{\dagger}\sbr{J^{\mu}_a\, T_a}{R_{\mu}}\,U
\\
&=& \ve_{abc}\,U^{\dagger}\,T_c\,U\left[ m_0^2\,e_0^2\,h_{ad}\,R^{d\,,\,\mu}\,R^b_{\mu}+
\frac{1}{2}h^{-1}_{bd}H^{d\,,\,\mu\nu}\,H^a_{\mu\nu}\right]
\nonumber
\er 
and that vanishes as a consequence of \rf{elforh}.

Note that the introduction of the fields $h_{ab}$ in the action \rf{model}, does not destroy the target space symmetries of  the original Skyrme model, namely $SU(2)_L\otimes SU(2)_R$. One still has six (left and right) conserved currents, and one can say, like in the original Skyrme model, that the equations of motion are equivalent to the conservation of these currents. Indeed, \rf{elforu} and \rf{elforh} follow from \rf{conscurr} and \rf{rightcurr}.  The introduction of the fields $h_{ab}$, however, brings two new structures to the  the theory \rf{model} as compared to the original Skyrme model. It allows the existence of an exact self-dual (BPS) sector, as we explain below in section \ref{subsec:bps}, and it renders the theory \rf{model} conformally invariant in the three dimensional space, as we explain in the appendix  \ref{app:conformal}. Indeed, one can check that the static energy associated to \rf{model}, given by 
\be
 E=\int d^3 x \left[ \frac{m_0^2}{2}\, h_{ab}\,R^a_{i}\,R^{b}_{i}+\frac{1}{4\,e_0^2}\, h^{-1}_{ab}\,H^a_{ij}\,H^{b}_{ij}\right]
 \lab{staticenergy}
\ee
is invariant under the conformal transformations in three dimensions (see \rf{densitiestransf}). In addition, the static version of the equations of motion \rf{elforu} and \rf{elforh}, are also conformally invariant (see \rf{eqofmotionconformal}). 

Note that by contracting the Euler-Lagrange equation \rf{elforh} with $h_{ab}$, one gets that the two terms of the Lagrangian density in \rf{model} must be equal on-shell, i.e.    
\be
\frac{m_0^2}{2}\, h_{ab}\,R^a_{\mu}\,R^{b\,,\, \mu}=-\frac{1}{4\,e_0^2}\, h^{-1}_{ab}\,H^a_{\mu\nu}\,H^{b\,,\,\mu\nu}
\ee 
For static configurations that implies that the two terms in the energy density in \rf{staticenergy} are also equal
\be
 \frac{m_0^2}{2}\, h_{ab}\,R^a_{i}\,R^{b}_{i}=\frac{1}{4\,e_0^2}\, h^{-1}_{ab}\,H^a_{ij}\,H^{b}_{ij}
 \lab{fakederrick}
 \ee
 In the original Skyrme model, where $h_{ab}$ is the identity matrix, a relation equivalent to 
 \rf{fakederrick} is obtained by Derrick's scaling argument \cite{derrick}.   The static energy \rf{staticenergy} is invariant under the scaling of the coordinates $x_i\rightarrow \alpha\, x_i$, due to the conformal symmetry. So, the static solutions of the theory \rf{model} do not have a fixed size, as the original Skyrmions do. However, the balance between the quadratic and quartic terms in space derivatives of the energy density of the theory \rf{model}, is provided by the Euler-Lagrange equations for the fields $h_{ab}$ as shown in \rf{fakederrick}. 
 
 \subsection{The static sector}
\label{subsec:static}
 
Another important role of the static Euler-Lagrange equations  for $h$ \rf{elforh} is to relate  $h$ to a real and symmetric matrix given by
\be
\tau_{ab}\equiv R_{i}^a\,R_{i}^b
\lab{ourstraintensor}
\ee
where the sum is over  the space index $i=1,2,3$, only. That matrix   is similar to the usual strain tensor of the Skyrme model defined by \cite{mantonstrain,mantonbook}
\be
D_{ij}=R_i^a\,R_j^a
\lab{mantonstrain}
\ee
Note that 
\be
{\rm det}\,\tau={\rm det}\,D\;;\qquad\qquad\qquad\qquad {\rm Tr}\,\tau^n={\rm Tr}\,D^n
\ee
and so the eigenvalues of $\tau$ and $D$ are the same. Now if $v_a$ is an arbitrary real vector then 
\be
v^T\, \tau\, v= \sum_{i=1}^3\(v_a\,R_i^a\)^2 \geq 0
\lab{positivetau}
\ee
Then $\tau$ is a positive semidefinite matrix and so its eigenvalues are all non-negative. At the same time the eigenvalues of the matrix $h$ can not vanish as it has to be invertible, and they have to be all positive for the energy of the theory \rf{model} to be positive definite. As a consequence of the static Euler-Lagrange equations  for $h$ \rf{elforh}, we then get the following results. Consider a domain ${\cal D}$ in the three dimensional space $\IR^3$. If the matrix $\tau$ is singular in ${\cal D}$, then the $SU(2)$ fields $U$ have to be constant in ${\cal D}$, i.e.
\be
{\rm det}\,\tau=0 \qquad\qquad \rightarrow \qquad\qquad U={\rm constant}\qquad\qquad {\rm for}\qquad x^i\in {\cal D}
\lab{uconstantresult}
\ee
and so the whole matrix $\tau$ vanishes (see \rf{rdef} and \rf{ourstraintensor}). 
On the other hand if none of the eigenvalues of $\tau$ vanish in ${\cal D}$, i.e. ${\rm det}\,\tau\neq 0$, then  it follows that the matrices $h$ and $\tau$  can be diagonalised by the same orthogonal matrix $M$ on that domain, i.e. 
\be
\(M^T\,h\,M\)_{ab}= \vp_a\,\delta_{ab}\;;\qquad\qquad \qquad \qquad \(M^T\,\tau\,M\)_{ab}= \omega_a\,\delta_{ab}
\lab{eigenvalueshtau}
\ee
and so they commute in ${\cal D}$, i.e. $\sbr{h}{\tau}=0$. 
In addition, their eigenvalues are related by 
\be
\omega_1=m_0^2\,e_0^2 \,\vp_2\,\vp_3\;;\qquad\qquad
\omega_2=m_0^2\,e_0^2 \,\vp_1\,\vp_3\;;\qquad\qquad 
\omega_3=m_0^2\,e_0^2 \,\vp_1\,\vp_2
\lab{eigenvalueshtaurel}
\ee
The proof of the relations \rf{uconstantresult}, \rf{eigenvalueshtau} and  \rf{eigenvalueshtaurel} is given in Appendix \ref{app:proof}. 

Note that the matrix $\tau$, defined in \rf{ourstraintensor}, transforms under the action of the group $SU(2)_L\otimes SU(2)_R$, given in \rf{leftsymm} and \rf{rightsymm}, in the same way as the matrix $h$. Therefore, the eigenvalues of $h$ and $\tau$ are invariant under the group $SU(2)_L\otimes SU(2)_R$. Indeed, if $\ket{v}$ is an eigenvector of $h$  with eigenvalue $\vp$, i.e. $h\ket{v}=\vp\ket{v}$, then $d\(g_L\)\ket{v}$ is an eigenvector of the transformed matrix $d\(g_L\)h\,d\(g_L\)^T$, with the same eigenvalue $\vp$. The same applies to the matrix $\tau$. 

The relations \rf{eigenvalueshtau} and  \rf{eigenvalueshtaurel} have some important consequences for the structure of the static solutions of the model \rf{model}. Using \rf{eigenvalueshtau} and  \rf{eigenvalueshtaurel} we can write that
\be
\(M^T\,\tau^{-1}\,M\)_{ab}=\frac{1}{m_0^2\,e_0^2 \,\vp_1\,\vp_2\,\vp_3}\; \vp_a\,\delta_{ab}
\ee
and so  the matrices $h$ and $\tau^{-1}$ are proportional in ${\cal D}$
\be
h=m_0^2\,e_0^2\, \({\rm det}\,h\)\; \tau^{-1}=\frac{\sqrt{{\rm det}\,\tau}}{\mid m_0\,e_0\mid}\; \tau^{-1}
\lab{goodrelhtau}
\ee
From \rf{rdef} and \rf{ourstraintensor} we observe that the matrix $\tau$ is a functional of the $SU(2)$  fields $U$ and their first derivatives only. Therefore, what the relation \rf{goodrelhtau} is telling us is that for static solutions of the theory \rf{model}, and in regions where the matrix $\tau$ is non-singular, the fields $h_{ab}$ are explicitly written in terms of the $U$ fields and their first derivatives without any integration needed. Note that this is a consequence of only the static Euler-Lagrange equations for the fields $h$ \rf{elforh}, and the positivity of the eigenvalues of $h$ and $\tau$. In the cases where the matrix $\tau$ is singular, it is also possible to find a matrix $h$ that solves the equations \rf{elforh} (see subsection \ref{subsec:tausing} of appendix \ref{app:proof}). We have shown that by using the Euler-Lagrange equations for $h$ \rf{elforh}, one can write the Euler-Lagrange equation for $U$ \rf{elforu} as in \rf{conscurr}. Then using \rf{goodrelhtau} one can express, in regions where $\tau$ is non-singular, the static version of \rf{conscurr} only in terms of the $U$ fields as 
\be
\partial_{i}J_{i}^a=0\;;\qquad \qquad {\rm with} \qquad \qquad 
J_{i}^a=\sqrt{{\rm det}\,\tau}\, \tau^{-1}_{ab}\,R_i^{b}+\frac{m_0^2\,e_0^2}{\sqrt{{\rm det}\,\tau}}\;\ve_{abc}\,R^b_{j}\,\tau_{cd}\,H^{d}_{ij}
\lab{conscurrstatic}
\ee

An interesting consequence of the equation \rf{goodrelhtau}  is that it implies that the determinant of the matrix $h_{ab}$ is proportional to the density of topological charge. In order to see that, let us treat the quantities $R_i^a$, introduced in \rf{rdef}, as a $3\times 3$ matrix with the following ordering of rows and columns $R_i^a\equiv \(R\)_{ia}$, $i=1,2,3$ and $a=1,2,3$.  We then have that
 \be
 \ve_{ijk}\,R_i^a\,R_j^b\,R_k^c=\ve_{abc}\,\ve_{ijk}\,R_{i1}\,R_{j2}\,R_{k3}=\ve_{abc}\,{\rm det} R
 \lab{nicedet}
 \ee
 Using \rf{su2killing} and \rf{nicedet} one then gets that
 \be
 \ve_{ijk}\,\trace\(R_i\,R_j\,R_k\)=i\, 3\, {\rm det} R
 \lab{detrtopcharge}
 \ee
 Then from \rf{topcharge} one gets 
 \be
 Q= -\frac{1}{16\,\pi^2}\int d^3x\; {\rm det} R
 \lab{topcharger}
 \ee
 From  \rf{goodrelhtau} it follows that, in regions where $\tau$ is non-singular,  ${\rm det}\,\tau= \(m_0^2\,e_0^2\)^3\,\({\rm det}\, h\)^2$, and \rf{ourstraintensor} implies ${\rm det}\,\tau=\({\rm det}\,R\)^2$. Therefore
 \be 
 {\rm det}\,h={\rm det}\,R/\lambda^3\qquad\qquad\qquad {\rm with}\qquad\qquad \lambda=\pm m_0\,e_0
 \lab{nicerhrel}
 \ee
 In order for the energy of the theory \rf{model} to be positive definite we need the eigenvalues of $h$ to be all positive, i.e.   ${\rm det}\,h>0$. Therefore, we conclude that
 \be
 {\rm sign}\(\lambda\, {\rm det}\,R\)=+1
 \lab{signlambdadetr}
 \ee
 and so ${\rm det}\,R$ has to have the same sign in all points in $\IR^3$. 
Therefore, in regions where $\tau$ is non-singular, from \rf{detrtopcharge} one gets that
\be
{\rm det}\,h=-\frac{i}{3\,\lambda^3}\,\ve_{ijk}\,\trace\(R_i\,R_j\,R_k\)
\lab{dethtopchargerel}
\ee
If $\tau$ is singular only in zero measure sets, it follows that the topological charge \rf{topcharge} is given by 
\be
Q=-\frac{\lambda^3}{16\,\pi^2}\,\int d^3x\, {\rm det}\,h
\lab{topchargedeth}
\ee
Since we need ${\rm det}\,h>0$,  for the energy of the theory \rf{model} to be positive definite, we need 
 \be
 {\rm sign}\(\lambda\,Q\)=-1
 \lab{signlambdaq}
 \ee

The relation \rf{goodrelhtau} between the matrices $h$ and $\tau$, valid in regions where $\tau$ is non-singular,  shows that the matrix $h$ does give a measure of the strain of the map from the compactfied three dimensional space $S^3$ to $SU(2)$.  
In addition, we can write the quadratic and quartic terms of the static energy \rf{staticenergy} as functions of these eigenvalues, Indeed, from \rf{fakederrick} and \rf{eigenvalueshtaurel} we get that 
\be
 h_{ab}\,R^a_{i}\,R^{b}_{i}=\frac{1}{2\,\lambda^2}\, h^{-1}_{ab}\,H^a_{ij}\,H^{b}_{ij}={\rm Tr}\(h\,\tau\)
 =3\,\lambda^2\,\vp_1\,\vp_2\,\vp_3
 \lab{bpsdensity}
  \ee
 Therefore, the static energy density in \rf{staticenergy} evaluated on a given static solution of the theory \rf{model}  is 
\be
{\cal E}_{{\rm static}}\equiv \frac{m_0^2}{2}\, h_{ab}\,R^a_{i}\,R^{b}_{i}+\frac{1}{4\,e_0^2}\, h^{-1}_{ab}\,H^a_{ij}\,H^{b}_{ij}=m_0^2\,\lambda^2\;3\,\vp_1\,\vp_2\,\vp_3=\frac{3\,m_0^2}{\mid\lambda\mid}\,\sqrt{\omega_1\,\omega_2\,\omega_3}
\ee
On the other hand, using \rf{hdef}, \rf{mantonstrain}, \rf{eigenvalueshtau} and \rf{eigenvalueshtaurel}, one gets that the static energy for the usual Skyrme, evaluated on a static solution of the theory \rf{model},  is given by
\br
{\cal E}_{{\rm Skyrme}}&\equiv& \frac{m_0^2}{2}\,R^a_{i}\,R^{a}_{i}+\frac{1}{4\,e_0^2}\, \,H^a_{ij}\,H^{a}_{ij}=\frac{m_0^2}{2}\,\left[{\rm Tr} D +\frac{1}{2\,\lambda^2}\(\({\rm Tr}D\)^2-{\rm Tr}\,D^2\)\right]
\nonumber\\
&=&\frac{m_0^2}{2}\,\left[\omega_1+\omega_2+\omega_3+\frac{1}{\lambda^2}\(\omega_1\,\omega_2+\omega_2\,\omega_3+\omega_1\,\omega_3\)\right]
\nonumber\\
&=&\frac{m_0^2\,\lambda^2}{2}\,\left[\vp_1\,\vp_2+\vp_2\,\vp_3+\vp_1\,\vp_3+\vp_1\,\vp_2\,\vp_3\,\(\vp_1+\vp_2+\vp_3\)\right]
\lab{skyrmedensity}
\er
where in the last equality we have assumed the relation \rf{eigenvalueshtaurel}, i.e. that the field configuration satisfy the static Euler-Lagrange equations for $h$ \rf{elforh}. Therefore, we can write that
\br
\Delta {\cal E}&\equiv& {\cal E}_{{\rm Skyrme}}-{\cal E}_{{\rm static}}
\nonumber\\
&=&\frac{m_0^2}{2}\,\left[
\(\sqrt{\omega_1}-\frac{\sqrt{\omega_2\,\omega_3}}{\mid\lambda\mid}\)^2
+\(\sqrt{\omega_2}-\frac{\sqrt{\omega_1\,\omega_3}}{\mid\lambda\mid}\)^2
+\(\sqrt{\omega_3}-\frac{\sqrt{\omega_1\,\omega_2}}{\mid\lambda\mid}\)^2
\right]
\nonumber\\
&=&\frac{m_0^2\,\lambda^2}{2}\,\left[\(1-\vp_1\)^2\,\vp_2\,\vp_3+
\(1-\vp_2\)^2\,\vp_1\,\vp_3+\(1-\vp_3\)^2\,\vp_1\,\vp_2\right]
\lab{deltaenergy}
\er
where again  in the last equality we have assumed the relation \rf{eigenvalueshtaurel}. 
Consequently, $\Delta {\cal E}$ is positive if the eigenvalues of the matrix $h$ are positive, and so if the energy of the theory \rf{model} is positive definite. On the domain where all $\vp_a$'s are positive we have three extrema for $\Delta{\cal E}$. The points $\vp_a=0$ and $\vp_a=1$, for $a=1,2,3$, are minima and the point $\vp_a=1/2$, $a=1,2,3$, is a saddle point.

\subsection{The self-dual sector}
\label{subsec:bps}

As explained in \cite{laf}, the introduction of the matrix $h_{ab}$ allows the existence of an exact self-dual sector, and makes the theory conformally invariant in three dimensional spatial sub manifold. The self-duality equations, defining that exact-self-dual sector are given by
\be
\lambda \,h_{ab}\,R^b_i=\frac{1}{2}\,\ve_{ijk}\,H^a_{jk}\;;\qquad\qquad\qquad \qquad\lambda=\pm\,m_0\,e_0
\lab{selfdualeqs}
\ee
Note that the self-duality equations \rf{selfdualeqs} can be written in $3$-vector space notation as 
\be
{\vec \nabla}\wedge {\vec R}_a =\lambda\, h_{ab}\,{\vec R}_b
\lab{genforcefree}
\ee
which is a generalization, to several vectors, of the well known force-free equation for magnetic fields in solar and plasma physics \cite{marsh,chandra,ua845,shnir}.   

The solutions of the first order equations \rf{selfdualeqs} not only solve  the static second order Euler-Lagrange equations \rf{elforu},   associated to the $SU(2)$ $U$-fields, but also solve those associated to the six extra fields $h_{ab}$, given in \rf{elforh}. Note that  the static version of \rf{elforh} can be written as
\be
\left[\mid \lambda\mid \,R_i^a-\frac{1}{2}\,h^{-1}_{ac}\,\ve_{ijk}\,H_{jk}^c\right]\,
\left[\mid \lambda\mid \,R_i^b+\frac{1}{2}\,h^{-1}_{bd}\,\ve_{ilm}\,H_{lm}^d\right]=0
\lab{factorstaticeqomh}
\ee
as the crossed terms cancel each other. Indeed, using \rf{nicedet} and \rf{hdef} it follows that, $h^{-1}_{ac}\,\ve_{ijk}\,H_{jk}^c\,R_i^b=h^{-1}_{ac}\,\ve_{cde}\,\ve_{ijk}\,R_i^b\,R_j^d\,R_k^e=2\, h^{-1}_{ab}\, {\rm det}\,R$. Similarly, $h^{-1}_{bd}\ve_{ilm}H_{lm}^d R_i^a=2h^{-1}_{ab} {\rm det}R$, and so they indeed cancel each other. Then \rf{factorstaticeqomh} is the same as the static version of \rf{elforh}, and \rf{selfdualeqs} does imply \rf{factorstaticeqomh}. 
 Using \rf{selfdualeqs} twice, one gets that 
 \be
 \ve_{abc}\,\lambda^2\, R_{i}^b\,h_{cd}\,R^{d}_{i}= \frac{1}{4}\, \ve_{abc}\,\ve_{ijk}\,h_{bd}^{-1}\,H^d_{jk}\,\ve_{ilm}\,H^c_{lm}=\frac{1}{2}\, \ve_{abc}\,h_{bd}^{-1}\,H^d_{jk}\,H^c_{jk}
 \ee
  and that implies that the static version of r.h.s. of \rf{elforu} vanishes. Now contract the static version of the l.h.s. of \rf{elforu} with $T_a$, and use \rf{selfdualeqs} to get 
\br
&\partial_i&\left[\lambda^2\, h_{ab}\,R^{b}_i+\ve_{abc}\,R^b_{j}\,h_{cd}^{-1}\,H^{d}_{ij}\right]\,T_a=\frac{\lambda}{2}\,\ve_{ijk}\,\partial_i\left[H^a_{jk}+2\,\ve_{abc}\,R^b_{j}\,R^c_k\right]\,T_a
\nonumber\\
&=&-i\,\frac{3\,\lambda}{2}\,\ve_{ijk}\,\partial_i \sbr{R_j}{R_k}=\frac{3\,\lambda}{2}\,\ve_{ijk}\,\sbr{\sbr{R_i}{R_j}}{R_k}=0
\er
where we have used the fact that $R_i$ satisfies the Maurer-Cartan equation \rf{maurercartaneq}, and in the last equality we have used the Jacobi identity. So, \rf{selfdualeqs} does imply the static versions of \rf{elforu} and \rf{elforh}.

The solutions of \rf{selfdualeqs}  saturate a Bogomolny-type bound of the static energy associated to \rf{model}.  Indeed, one can write the  static energy  \rf{staticenergy} as
 \br
E =\frac{1}{2\,e_0^2}\,\int d^3 x \left[ \lambda\, R_i^b\,k_{ba}-\frac{1}{2}\,k^{-1}_{ab}\,\ve_{ijk}\,H_{jk}^b\right]^2-{\rm sign}\(\lambda\)\,\frac{\mid m_0\mid}{\mid e_0\mid}\,48\,\pi^2\,Q
\lab{prebpsenergy}
 \er
where $Q$ is the topological charge, given in \rf{topcharge},  and where we have written the matrix $h$ in terms of a real and invertible matrix $k$ as 
 \be
 h=k\,k^T
 \lab{haskk}
 \ee
 Note that if $O$ is an orthogonal $3\times 3$ matrix ($O\,O^T=\one$), then $k$ and $k\,O$ give the same $h$. Such a freedom accounts for the three extra entries of $k$ as compared with those of $h$, since $O$ has three independent entries. In fact, the theory \rf{model} was constructed in \cite{laf} using arguments of \cite{selfdual} based on a splitting of the density of the topological charge \rf{topcharge}. That splitting procedure implies that the matrix $h$ has the form \rf{haskk}.     Therefore, using \rf{signlambdaq} we get that the self-dual solutions saturate the bound of the static energy, and in that case it is given by
 \br
 E_{{\rm BPS}}=m_0^2\, \int d^3 x \,  h_{ab}\,R^a_{i}\,R^{b}_{i}=
\frac{1}{2\,e_0^2}\,\int d^3 x\,  h^{-1}_{ab}\,H^a_{ij}\,H^{b}_{ij}=\frac{\mid m_0\mid}{\mid e_0\mid}\,48\,\pi^2\,\mid Q\mid
\lab{bpsenergy}
 \er
 We now want to show that the static Euler-Lagrange equations for the fields $h_{ab}$ \rf{elforh} imply the self-duality equations \rf{selfdualeqs}. Let us denote
 \be
 S_i^{(\pm),a}\equiv \mid \lambda\mid \,R_i^a\pm \frac{1}{2}\,h^{-1}_{ac}\,\ve_{ijk}\,H_{jk}^c
 \lab{siapmdef}
 \ee
 Then the static Euler-Lagrange equations for the fields $h_{ab}$, as written in \rf{factorstaticeqomh}, can be expressed as
 \be
 S_i^{(+),a}\,S_i^{(-),b}=0
 \lab{factorstaticeqomhfors}
 \ee
 From \rf{siapmdef} we have
 \be
 S_i^{(+),a}+S_i^{(-),a}= 2\, \mid \lambda\mid \,R_i^a
 \lab{sumsiapm}
 \ee
Contracting \rf{sumsiapm}  with  $S_i^{(\pm),b}$ and using \rf{factorstaticeqomhfors} one gets
\br
S_i^{(\pm),a}\,S_i^{(\pm),b}&=&2\, \mid \lambda\mid \,R_i^a\, S_i^{(\pm),b}=2\,\mid \lambda\mid\left[\mid \lambda\mid\,\tau_{ab}\pm \({\rm det}\,R\)\,h^{-1}_{ab}\right]
\nonumber\\
&=&2\,\mid \lambda\mid^2\left[\tau_{ab}\pm \({\rm sign}\,\lambda\)\,\mid \lambda\mid^2\,\({\rm det}\,h\)\,h^{-1}_{ab}\right]
\nonumber\\
&=&2\,\mid \lambda\mid^2\,\tau_{ab}\left[1\pm \({\rm sign}\,\lambda\)\right]
 \er
 where we have used \rf{ourstraintensor}, \rf{nicedet}, \rf{nicerhrel} and \rf{goodrelhtau}. Consequently
 \br
\qquad\qquad {\rm sign}\,\lambda =\pm 1\qquad\qquad
 \rightarrow \qquad\qquad S_i^{(\mp),a}=0
 \er
 But the vanishing of $S_i^{(\pm),a}$, as given in \rf{siapmdef}, is equivalent to the self-duality equations \rf{selfdualeqs}. Therefore, we come to a very important conclusion: the static Euler-Lagrange equations for the fields $h_{ab}$ \rf{elforh} imply the self-duality equations \rf{selfdualeqs}, and these in their turn imply the static Euler-Lagrange equations for the $SU(2)$ fields $U$ \rf{elforu}. So, the static and self-dual sectors of the theory \rf{model} coincide. Consequently, the only non-self-dual solutions of the theory \rf{model} must necessarily be time dependent. 
 
 Note in addition that in the relation \rf{goodrelhtau}, which is a direct consequence of the static Euler-Lagrange equations for the fields $h_{ab}$ \rf{elforh} , the matrix $h$ is expressed entirely in terms of the matrix $\tau$,  which in its turn is expressed in terms of the $SU(2)$ fields $U$ and their first derivatives. Therefore, by choosing any static field configuration for the $U$-fields one gets from \rf{goodrelhtau} the matrix $h$ that satisfies the static Euler-Lagrange equations for the fields $h_{ab}$ \rf{elforh}, and so both $U$ and $h$ are static solutions of the theory \rf{model}. Note that in the cases where the matrix $\tau$ is singular, the relation \rf{goodrelhtau} does not hold true. However, as we show in subsection \ref{subsec:tausing} of appendix \ref{app:proof}, it is possible to find a matrix $h$ that solves the self-duality equations \rf{selfdualeqs} even when $\tau$ is singular. Consequently, \rf{model} has plenty of static solutions. We now analyse some special types of these static solutions.

\section{The Holomorphic Ansatz}
\label{sec:ansatzholom}
\setcounter{equation}{0}

In this section we construct exact self-dual Skyrmion solutions for the self-duality equations \rf{selfdualeqs} using the so-called rational map ansatz \cite{rational1,rational2,mantonbook}.  We parameterize the $SU(2)$ group elements $U$, with a real scalar field $f$ and a complex scalar field $u$, together with its complex conjugate $\ub$,  as \cite{skyrmejoaq,laf}
\be
U= W^{\dagger}\, e^{i\,f\, T_3}\,W
\lab{udecompdef}
\ee
with $W$ having the following form in the spinor representation of $SU(2)$
\br
W=\frac{1}{\sqrt{1+\u2}}\(
\begin{array}{cc}
1& i\,u\\
i\, \ub&1
\end{array}\)
\er
Therefore we have that
\be
R_{i} = i\,\partial_{i}U\,U^{\dagger}=- V\,\Sigma_i\,V^{\dagger}\;;\qquad\qquad\qquad\qquad 
V\equiv W^{\dagger}\,e^{i\,f\,T_3/2}
\ee
and
\be
\Sigma_i\equiv \partial_if\,T_3+\frac{i\,2\,\sin\(f/2\)}{1+\u2}\(\partial_i u\,T_{+}-\partial_i\ub\,T_{-}\)
\ee
with $T_{\pm}=T_1\pm i\,T_2$. The self-duality equations \rf{selfdualeqs} become
\be
\lambda\, {\rm Tr}\( \Sigma_i\, T_b\){\tilde h}_{ba}=\frac{i}{2}\,\ve_{ijk}\,{\rm Tr}\(\sbr{\Sigma_j}{\Sigma_k}\,T_a\)
\lab{selfdualeqssigma}
\ee
where we have introduced the matrix ${\tilde h}_{ab}$ as 
\be
h_{ab}\equiv d_{ac}\(V\)\,{\tilde h}_{cd}\,d^T_{db}\(V\)
\lab{htildedef}
\ee
with $d\(V\)$ being the adjoint representation matrix for the group element $V=W^{\dagger}\,e^{i\,f\,T_3/2}$ (see \rf{adjoint}). The adjoint representation (triplet) is real and unitary and so  $d$ is a $3\times 3$ orthogonal matrix. Therefore, since $h$ is symmetric, so is the matrix  ${\tilde h}$. 

We now use spherical coordinates, but instead of using the polar and azimuthal angles we stereographic project the two sphere on a plane and parameterize that plane by a complex coordinate $w$, together with its complex conjugate $\wb$. So, we have the coordinate transformation
\be
x_1=r\;\frac{-i\(w-\wb\)}{1+\w2}\;;\qquad \qquad x_2=r\;\frac{\(w+\wb\)}{1+\w2}\;;\qquad \qquad
x_3=r\;\frac{\w2-1}{1+\w2}
\lab{wcartesiandef}
\ee
where $r$ is the radial distance. The  Euclidean space metric becomes 
\be
ds^2=dr^2+\frac{4\,r^2}{\(1+\w2\)^2}\,dw\,d\wb
\lab{holometric}
\ee
The self-duality equations \rf{selfdualeqssigma} become
\br
\lambda\, {\rm Tr}\( \Sigma_r\, T_b\){\tilde h}_{ba}&=&
\frac{\(1+\w2\)^2}{2\,r^2}\,{\rm Tr}\(\sbr{\Sigma_w}{\Sigma_{\wb}}\,T_a\)
\nonumber\\
\lambda\, {\rm Tr}\( \Sigma_{\wb}\, T_b\){\tilde h}_{ba}&=&
{\rm Tr}\(\sbr{\Sigma_{\wb}}{\Sigma_{r}}\,T_a\)
\lab{selfdualeqsrzzb}\\
\lambda\, {\rm Tr}\( \Sigma_{w}\, T_b\){\tilde h}_{ba}&=&
{\rm Tr}\(\sbr{\Sigma_{r}}{\Sigma_{w}}\,T_a\)
\nonumber
\er
We now use the holomorphic ansatz 
\be
f\equiv f\(r\)\;;\qquad\qquad \qquad u\equiv u\(w\)\;;\qquad\qquad \qquad \ub\equiv \ub\(\wb\)
\lab{holoansatz}
\ee
and so
\be
\Sigma_r=\partial_r f\,T_3\;;\qquad\qquad 
\Sigma_w=\frac{i\,2\,\sin\(f/2\)}{1+\u2}\,\partial_w u\,T_{+}\;;\qquad\qquad 
\Sigma_{\wb}=-\frac{i\,2\,\sin\(f/2\)}{1+\u2}\,\partial_{\wb} \ub\,T_{-}
\ee
Therefore, we get from \rf{selfdualeqsrzzb} that ${\tilde h}_{ab}$ is diagonal
\br
{\tilde h}_{11}={\tilde h}_{22}=\frac{\partial_rf}{\lambda}\;;\qquad
{\tilde h}_{33}=\frac{4\,\sin^2\(f/2\)}{\lambda\,r^2\,\partial_r f}\,\frac{\(1+\w2\)^2}{\(1+\u2\)^2}\,\partial_w u\,\partial_{\wb}\ub\;;\qquad {\tilde h}_{12}={\tilde h}_{13}={\tilde h}_{23}=0
\nonumber\\
\lab{htildeholo}
\er
Consequently, in order to have the eigenvalues of the matrix $h$ positive, we have to impose that (see \rf{signlambdaq})
\be
{\rm sign}\(\lambda\,\partial_rf\)=+1
\lab{signlambdaf}
\ee
and so, $f$ has to be a monotonic function of $r$, monotonically increasing for $\lambda>0$ and monotonically decreasing for $\lambda<0$.

The matrix $h$ can be obtained from \rf{htildedef} with $d\(V\)=d\(W^{\dagger}\)\,d\(e^{i\,f\,T_3/2}\)$, and 
\br
d\(e^{i\,f\,T_3/2}\)=\left(
\begin{array}{ccc}
 \cos \frac{f}{2} & \sin \frac{f}{2} & 0 \\
 -\sin \frac{f}{2} & \cos \frac{f}{2} & 0 \\
 0 & 0 & 1 \\
\end{array}
\right)
\er
and
\br
d\(W^{\dagger}\)=\frac{1}{1+\u2}\,\left(
\begin{array}{ccc}
 \frac{1}{2} \left(2+u^2+\ub^2\right) & \frac{1}{2} i \left(u^2-\ub^2\right) & i (u-\ub) \\
 \frac{1}{2} i \left(u^2-\ub^2\right) & \frac{1}{2} \left(2-u^2-\ub^2\right) & -(u+\ub ) \\
 -i (u-\ub) & u+\ub & 1-\u2  \\
\end{array}
\right)
\er
We then have that
\be
{\rm det}\,h={\rm det}\,{\tilde h}= 
\frac{4}{\lambda^3}\,\frac{\partial_r f\,\sin^2\(f/2\)}{r^2}\,\frac{\(1+\w2\)^2}{\(1+\u2\)^2}\,\partial_w u\,\partial_{\wb}\ub
\ee
Therefore, from \rf{topchargedeth} we have that the topological charges of these configurations are
\be
Q=\frac{i}{4\,\pi^2}\int dw\wedge d\wb\,\frac{\partial_w u\,\partial_{\wb}\ub}{\(1+\u2\)^2}\;\left[f-\sin f\right]\mid_{r=0}^{r=\infty}
\lab{pretopcharge}
\ee
where we have used the fact that the volume element is $d^3 x=-\frac{i\,2\,r^2}{\(1+\w2\)^2}\,dr\wedge dw\wedge d\wb$ (see \rf{holometric}). 

Note that the self-duality equations \rf{selfdualeqs} do not determine the functions $f\(r\)$, $u\( w\)$ and $\ub\(\wb\)$. It only determines the matrix $h$ in terms of these functions. Therefore, any holomorphic ansatz configuration \rf{holoansatz} leads to a solution of the self-duality equations \rf{selfdualeqs}.  However, for the function $u\(w\)$ to be a well defined map between two-spheres it has to be a ratio of two polynomials $p$ and $q$, i.e. the so-called rational map
\be
u\(w\)=\frac{p\(w\)}{q\(w\)}
\lab{rationalmap}
\ee
The degree $n$ of the map between two-spheres defined by \rf{rationalmap} is the highest power of $w$ in either of the polynomials $p$ or $q$, and it is equal to \cite{mantonbook} 
\be
n=\frac{i}{2\,\pi}\int dw\wedge d\wb\,\frac{\mid\,q\,\partial_w p-p\,\partial_w q\mid^2}{\(\mid p\mid^2+\mid q\mid^2\)^2}
\ee
In addition, for the topological charge \rf{pretopcharge} to be non-trivial the profile function $f\(r\)$ has to be such that the quantity $\left[f-\sin f\right]\mid_{r=0}^{r=\infty}$ does not vanish.  For a given integer $m$  one has that $e^{i\,2\,\pi\,m\,T_3}=\pm \one$, depending if the representation used is of integer (+) or half-integer (-) spin. Therefore, from \rf{udecompdef} we have that if $f=2\,\pi\,m$ then $U=\pm \one$. Consequently, we shall consider boundary conditions such that $f\(0\)= 2\,\pi\,m$, and $f\(\infty\)=0$, and so $U\(\infty\)=\one$ and $U\(0\)=\pm \one$, depending if the representation used for $U$ is of integer (+) or half-integer (-) spin. The topological charge \rf{pretopcharge} then becomes
\be
Q=-m\,n
\ee
Clearly, if we swap the boundary conditions for $f$ at the origin and at spatial infinity, the topological charge changes sign. 

Note that alternatively we could have chosen the boundary condition for $f$ such that $f\(0\)= 2\,\pi$, and $f\(\infty\)=0$, leading to a solution of topological charge $(-n)$. We could then use  the product ansatz  \cite{mantonbook} to construct a  Skyrmion of charge $(-m\,n)$ by taking the $SU(2)$ field $U$ as  $U_{(-m\,n)}= U_{(-n)}^m$, where $U_{(-n)}$ is the field $U$ for the Skyrmion of charge $(-n)$, i.e. using  \rf{udecompdef} one has  $U_{(-n)}= W^{\dagger}\, e^{i\,f_{(1)}\, T_3}\,W$, with  $f_{(1)}\(0\)= 2\,\pi$, and $f_{(1)}\(\infty\)=0$. Therefore,  $U_{(-m\,n)}= W^{\dagger}\, e^{i\,m\,f_{(1)}\, T_3}\,W$.  Consequently, for a given profile function $f_{(1)}$, the number of ways one can construct  a self-dual solution of topological charge $N$ is given by the number of partitions of $N$ into the product of two integers. As we have said before, any configuration for the $SU(2)$ field $U$ leads to a self-dual solution, since the $h$-fields act as spectators (see \rf{goodrelhtau}). So, one can change the profile function smoothly without changing its boundary values and the self-dual solution obtained has the same topological charge. So, the self-dual solutions are infinitely degenerated. 

In order to illustrate the kind of solutions we get for the matrix $h$ we  shall consider the simple profile function 
\be
f\(r\)=4\,{\rm ArcTan}\(\frac{1}{\zeta}\)\qquad\qquad\qquad \qquad \zeta=\frac{r}{a}
\lab{chosenf}
\ee
which implies 
\be
\frac{\partial_rf}{\lambda}=
\frac{4\,\sin^2\(f/2\)}{\lambda\,r^2\,\partial_r f}=-\frac{1}{\lambda\, a}\,\frac{4}{\(1+\zeta^2\)}
\ee
with $a$ being an arbitrary positive parameter of dimension of length. Then, from \rf{htildeholo}, we get that the eigenvalues of the matrix $h$ are
\be
\(\vp_1\,,\,\vp_2\,,\,\vp_3\)=\frac{1}{\mid\lambda\mid\, a}\,\frac{4}{\(1+\zeta^2\)}\(1\,,\,1\,,\, 
\frac{\(1+\w2\)^2}{\(1+\u2\)^2}\,\partial_w u\,\partial_{\wb}\ub\)
\lab{eigenvaluesconfig}
\ee
where we have used \rf{signlambdaf} and the fact that the profile function \rf{chosenf} is monotonically decreasing.

\section{The Toroidal Ansatz}
\label{sec:ansatztoro}
\setcounter{equation}{0}

It was shown in \cite{laf} (see \rf{trickone}) that the self-duality equations \rf{selfdualeqs} are invariant under conformal transformations on the three dimensional spatial sub manifold. We now use  that symmetry together with the target space symmetries \rf{leftsymm} and \rf{rightsymm} to build an ansatz based on the toroidal coordinates \cite{babelon,shnir}. In order to implement that we parameterize the $SU(2)$ group elements as 
\br
U= \(
\begin{array}{cc}
Z_2&i\,Z_1\\
i\,{\bar Z}_1& {\bar Z}_2
\end{array}\)\;;\qquad\qquad\qquad \mid Z_1\mid^2+\mid Z_2\mid^2=1
\lab{torou}
\er
We now select two commuting $U(1)$ subgroups of the target space symmetry group $SU(2)_L\otimes SU(2)_R$, given in \rf{leftsymm} and \rf{rightsymm}, as follows
\br
U\rightarrow e^{i\,\alpha\,T_3}\,U\,e^{-i\,\alpha\,T_3}\;;\qquad\qquad {\rm or}\qquad \qquad
Z_1\rightarrow e^{i\,\alpha}\,Z_1\;;\qquad Z_2\rightarrow Z_2
\lab{firstu1}
\er
and
\br
U\rightarrow e^{i\,\beta\,T_3}\,U\,e^{i\,\beta\,T_3}\;;\qquad\qquad {\rm or}\qquad \qquad
Z_1\rightarrow Z_1\;;\qquad Z_2\rightarrow e^{i\,\beta}\,Z_2
\lab{secondu1}
\er
We also select two commuting $U(1)$ subgroups of the conformal group in three dimensions defined by the vector fields \cite{babelon,shnir}
\br
\partial_{\phi}&\equiv& x^2\,\partial_1-x^1\,\partial_2
\lab{phitransf}
\\
\partial_{\xi}&\equiv&\frac{x^3}{a}\(x^1\,\partial_1+x^2\,\partial_2\)+\frac{1}{2\,a}\(a^2+x_3^2-x_1^2-x_2^2\)\,\partial_3
\lab{xitransf}
\er
where $a$ is a free  parameter with dimension of length, and where we have introduced the angles $\phi$ and $\xi$, such that the vector fields above generate rotations along these angular directions. The transformation \rf{phitransf} corresponds to rotations on the plane $x_1\,x_2$, and \rf{xitransf} to a linear combination of a special conformal transformation and a translation along the $x_3$ direction. It turns out that $\phi$ and $\xi$ correspond in fact to the two angles of the toroidal coordinates in three dimensions defined by
\br
x^1&=& \frac{a}{p}\,\sqrt{z}\,\cos\phi\;; \qquad\qquad\qquad \qquad p=1-\sqrt{1-z}\,\cos\xi
\nonumber\\
x^2&=& \frac{a}{p}\,\sqrt{z}\,\sin\phi\;; \qquad\qquad\qquad \qquad 0\leq z\leq 1
\lab{torocoord}\\
x^3&=& \frac{a}{p}\,\sqrt{1-z}\,\sin \xi\;; \qquad\qquad\qquad\;\; 0\leq \phi\,,\,\xi\leq 2\,\pi
\nonumber
\er
where the Euclidean metric becomes 
\be
ds^2=\frac{a^2}{p^2}\left[\frac{dz^2}{4\,z\(1-z\)}+\(1-z\)\,d\xi^2+z\,d\phi^2\right]
\lab{torometric}
\ee
We build an ansatz that is invariant under the joint action of the  $U(1)$  subgroups \rf{firstu1} and \rf{phitransf},  and also invariant under the joint action of the subgroups \rf{secondu1} and \rf{xitransf}, leading to 
\be
Z_1=\sqrt{F\( z\)}\,e^{i\,n\,\phi}\;;\qquad\qquad\qquad Z_2=\sqrt{1-F\( z\)}\,e^{i\,m\,\xi}\;;\qquad\qquad m\,,\,n \in \IZ
\lab{toroansatz}
\ee
with $0\leq F\leq 1$. From \rf{torou} and \rf{toroansatz} we get that 
\br
R_z&=&i\,\partial_zU\,U^{\dagger}=\frac{F^{\prime}}{\sqrt{F\(1-F\)}}\,\left[-\cos\(m\xi+n\phi\)\,T_1+\sin\(m\xi+n\phi\)\,T_2\right]
\lab{toror}\\
R_{\xi}&=&i\,\partial_{\xi}U\,U^{\dagger}=-2m\left[\sqrt{F\(1-F\)}\left[\sin\(m\xi+n\phi\)\,T_1+\cos\(m\xi+n\phi\)\,T_2\right]+\(1-F\)\,T_3\right]
\nonumber
\\
R_{\phi}&=&i\,\partial_{\phi}U\,U^{\dagger}=2n\left[\sqrt{F\(1-F\)}\left[\sin\(m\xi+n\phi\)\,T_1+\cos\(m\xi+n\phi\)\,T_2\right]-F\,T_3\right]
\nonumber
\er
and these quantities satisfy the commutation relations
\br
\sbr{R_z}{R_{\xi}}&=&-i\,\frac{m}{n}\,\frac{F^{\prime}}{F}\, R_{\phi}
\nonumber\\
\sbr{R_{\phi}}{R_{z}}&=&-i\,\frac{n}{m}\,\frac{F^{\prime}}{\(1-F\)}\, R_{\xi}
\lab{rcommrel}\\
\sbr{R_{\xi}}{R_{\phi}}&=&-i\,4\,m\,n\,\frac{F\(1-F\)}{F^{\prime}}\, R_{z}
\nonumber
\er
Then the  self-duality equations \rf{selfdualeqs} become
\br
\lambda\,\frac{a}{p}\, R_z^b\,h_{ba}&=&-2\,m\,n\,\frac{1}{F^{\prime}}\,\frac{F\(1-F\)}{z\(1-z\)}\, R_z^a
\nonumber\\
\lambda\,\frac{a}{p}\, R_{\xi}^b\,h_{ba}&=&-2\,\frac{n}{m}\, F^{\prime}\,\frac{\(1-z\)}{\(1-F\)}\, R_{\xi}^a
\lab{toroselfdualeqs}\\
\lambda\,\frac{a}{p}\, R_{\phi}^b\,h_{ba}&=&-2\,\frac{m}{n}\, F^{\prime}\,\frac{z}{F}\, R_{\phi}^a
\nonumber
\er
So, the self-duality equations \rf{toroselfdualeqs} imply that the three  quantities $R_z^a$, $R_{\xi}^a$ and $R_{\phi}^a$, are eigenvectors of the matrix $h_{ab}$. Since eigenvectors and eigenvalues are known, one can reconstruct that matrix, obtaining
\be
h= M\, h_D\, M^T\;;\qquad\qquad\qquad\qquad\qquad M\,M^T=\one
\lab{hdiagonaltoroidaldef}
\ee
with
\be
h_D=-\frac{2}{\lambda}\,\frac{p}{a}\, 
\mbox{\rm diag.}\,\left(m\,n\,\frac{1}{F^{\prime}}\,\frac{F\(1-F\)}{z\(1-z\)}\;,\; \frac{n}{m}\,
F^{\prime}\,\frac{\(1-z\)}{\(1-F\)}\;,\; \frac{m}{n}\,F^{\prime}\,\frac{z}{F}\right)
\equiv \mbox{\rm diag.}\,\left(\vp_1\,,\,\vp_2\,,\,\vp_3\right)
\lab{hdiagonaltoroidalfinal}
\ee
and
\br
M=\(
\begin{array}{ccc}
\cos\(m\xi+n\phi\)&\sqrt{F}\,\sin\(m\xi+n\phi\)&\sqrt{1-F}\,\sin\(m\xi+n\phi\)\\
-\sin\(m\xi+n\phi\)&\sqrt{F}\,\cos\(m\xi+n\phi\)&\sqrt{1-F}\,\cos\(m\xi+n\phi\)\\
0&\sqrt{1-F}&-\sqrt{F}
\end{array}\)
\lab{mtoroidaldef}
\er
Therefore
\be
{\rm det}\,h={\rm det}\,h^{({\rm diag.})}=-m\,n\,\left[\frac{2}{\lambda}\,\frac{p}{a}\right]^3\, F^{\prime}
\ee
Consequently, from \rf{topchargedeth} one gets that the topological charges for these configurations are
\be
Q=m\,n\,\left[ F\(1\)-F\(0\)\right]
\ee
where we have used the fact that the volume element is $d^3x=\frac{1}{2}\,\frac{a^3}{p^3}\,dz\,d\xi\,d\phi$ (see \rf{torometric}). 

Note that the self-duality equations do not determine the profile function $F\(z\)$. However, for the matrix $h$ to be invertible, one has to have $m , n \neq 0$, and $F^{\prime}\neq 0$. Note in addition that in order for the eigenvalues $\vp_a$ to be positive, one observes from \rf{hdiagonaltoroidalfinal}, that one must have
\be
{\rm sign}\(\lambda\,m\,n\,F^{\prime}\)=-1
\ee
Consequently, $F$ has to be a monotonic function of $z$, increasing for $\lambda\,m\,n<0$, and decreasing for $\lambda\,m\,n>0$. That is the equivalent of relation \rf{signlambdaf} in the holomorphic ansatz. 

The $\tau$-matrix, defined in \rf{ourstraintensor}, in the case of this toroidal ansatz, is given by
\be
\tau_{ab}=\frac{p^2}{a^2}\left[ 4\,z\,\(1-z\)\, R_z^a\,R_z^b+\frac{1}{1-z}\,R_{\xi}^a\,R_{\xi}^b+\frac{1}{z}\, R_{\phi}^a\,R_{\phi}^b\right]
\lab{tautoroidal}
\ee
and one can write it as
\be
\tau=M\,\tau_D\,M^T
\lab{diagonalizetautoroidal}
\ee
with $M$ given by \rf{mtoroidaldef}, and 
\be
\tau_D=4\, \frac{p^2}{a^2}\,{\rm diag.}\,\({F^{\prime}}^2\,\frac{z\(1-z\)}{F\(1-F\)}\,,\, m^2\,\frac{\(1-F\)}{\(1-z\)}\,,\, n^2\, \frac{F}{z}
\)\equiv {\rm diag.}\,\(\omega_1\,,\,\omega_2\,,\, \omega_3\)
\lab{tautoroidaldiag}
\ee
From \rf{hdiagonaltoroidaldef} and \rf{diagonalizetautoroidal} one sees that the matrices $h$ and $\tau$ are indeed diagonalised by the same orthogonal matrix $M$, given in \rf{mtoroidaldef}, and such a fact is compatible with \rf{eigenvalueshtau}. In addition, comparing  \rf{hdiagonaltoroidalfinal} and  \rf{tautoroidaldiag}  one observes their  eigenvalues  do satisfy the relations \rf{eigenvalueshtaurel}.

\section{Conclusions}
\label{sec:conclusion}
\setcounter{equation}{0}

We have studied the properties of a modified Skyrme model, originally proposed in \cite{laf}, that possesses an exact self-dual sector. The novelty of such a modification is the introduction of six scalar fields assembled into a symmetric and invertible matrix $h_{ab}$, that not only makes the existence of the self-dual sector possible, but also renders it conformally invariant in the spatial sub-manifold $\IR^3$. We have shown that the static and self-dual sector  are in fact equivalent, in the sense that any static solution also satisfies the self-duality equations. In addition, the fields $h_{ab}$ are spectators in the self-dual sector since  for any configuration for the $SU(2)$ $U$-fields, the fields $h_{ab}$ adjust themselves to satisfy the self-duality equations. Consequently, the model possesses an infinity of analytical self-dual solutions. We have construct explicitly two classes of such solutions: one based on the holomorphic rational ansatz, and another based on a toroidal ansatz constructed from the conformal symmetry of the  self-dual sector.  

The construction of the self-dual sector is such that it does not leave room for a kinetic term for the fields $h_{ab}$.  The addition of a kinetic term, as well as of a mass and potential terms, break the conformal symmetry of the theory in $\IR^3$, and also destroys the self-dual sector. However, that is desirable for physical applications, since the exact self-dual solutions have their energy proportional to the topological charge and so they do not have an interaction when at rest relative to each other.  We have verified that the addition of the kinetic, mass and potential terms can lead to a positive binding energy in some range of the coupling constants. That can be useful in application of the model to nuclear physics. The numerical results we have obtained on those lines will be presented elsewhere \cite{tocome}.

\vspace{2cm} 

\noindent {\bf Acknowledgements:} The authors are grateful to Carlos Naya for many helpful discussions. LAF is partially supported by Conselho Nacional de Desenvolvimento Cient\'ifico e Tecnol\'ogico - CNPq (contract 308894/2018-9) and Funda\c c\~ao de Amparo \`a Pesquisa do Estado de S\~ao Paulo - FAPESP (contract 2018/01290-6). LRL is supported by CAPES.  

\vspace{2cm} 

\appendix

\section{The conformal symmetry}
\label{app:conformal}
\setcounter{equation}{0}

Following \cite{laf,babelon} we introduce the conformal transformations in the Euclidean three dimensional space   ($i,j,k=1,2,3$)
\be
\delta x_i=\zeta_i\;; \qquad \qquad \qquad \qquad \qquad \partial_i\zeta_j+\partial_j\zeta_i=2\,D\,\delta_{ij} \qquad\quad
\lab{conformaltransf}
\ee
where the function $D$ vanishes for translations and rotations, it is constant for dilatations, and it is linear in the $x_i$'s for the special conformal transformations \cite{babelon}.  We take the fields to transform as 
\be
\delta U=0\;;\qquad\qquad \delta h_{ab}=-D\,h_{ab}\;;\qquad\qquad \delta h^{-1}_{ab}=D\,h^{-1}_{ab}
\lab{fieldtransf1}
\ee
It then follows that
\br
\delta R_i^a&=&-\partial_i\zeta_j\,R_j^a\;;\qquad\qquad\qquad 
\delta H_{ij}^a=-\partial_i\zeta_k\, H_{kj}^a-\partial_j\zeta_k\, H_{ik}^a
\lab{fieldtransf2}
\er
and
\br
\delta\(\partial_ih_{ab}\)&=&-\partial_i\zeta_j\,\partial_jh_{ab}-D\,\partial_ih_{ab}-h_{ab}\,\partial_iD\;;
\nonumber\\
\delta\(\partial_ih^{-1}_{ab}\)&=&-\partial_i\zeta_j\,\partial_jh^{-1}_{ab}+D\,\partial_ih^{-1}_{ab}+h^{-1}_{ab}\,\partial_iD
\lab{fieldtransf3}
\er
In addition we have that
\be
\delta\(\partial_iR_j^a\)=-\partial_i\zeta_k\,\partial_kR_j^a-R_k^a\partial_i\partial_j\zeta_k-
\partial_j\zeta_k\,\partial_iR_k^a
\lab{fieldtransf4}
\ee
and 
\br
\delta\(\partial_kH^a_{ij}\)&=&-\partial_k\zeta_l\,\partial_lH^a_{ij}-\partial_i\zeta_l\,\partial_kH^a_{lj}
-\partial_j\zeta_l\,\partial_kH^a_{il}-\partial_k\partial_i\zeta_l\, H^a_{lj}-\partial_k\partial_j\zeta_l\, H^a_{il}
\lab{fieldtransf5}
\er
Using \rf{conformaltransf}, \rf{fieldtransf1} and \rf{fieldtransf2} one gets that
\br
\delta\(h_{ab}\,R^a_i\,R^b_i\)&=&-3\,D\, h_{ab}\,R^a_i\,R^b_i
\nonumber\\
\delta\(h^{-1}_{ab}\,H^a_{ij}\,H^b_{ij}\)&=&-3\,D\, h^{-1}_{ab}\,H^a_{ij}\,H^b_{ij}
\lab{densitiestransf}\\
\delta\(\ve_{ijk}\,\ve_{abc}\,R^a_i\,R^b_j\,R^c_k\)&=&-3\,D\,\ve_{ijk}\,\ve_{abc}\,R^a_i\,R^b_j\,R^c_k
\nonumber
\er
The volume element transforms, under \rf{conformaltransf}, as $\delta\(d^3x\)=3\,D\, d^3x$, and so the static energy \rf{staticenergy}, and the topological charge \rf{topcharge}, are conformally invariant. 

We shall denote 
\br
\Lambda^{(1)}_a&\equiv& \partial_i\left[m_0^2\,e_0^2\,h_{ab}\,R^b_i+\ve_{abc}\,R^b_j\,h^{-1}_{cd}\,H^d_{ij}\right]
\nonumber\\
\Lambda^{(2)}_{ab}&\equiv&m_0^2\,e_0^2\,R^a_i\,R^b_i-\frac{1}{2}\,h^{-1}_{ac}\,H^c_{ij}\,h^{-1}_{bd}\,H^d_{ij}
\\
\Lambda^{(3)}_{a\,i}&\equiv&\lambda\,h_{ab}\,R_i^b-\frac{1}{2}\,\ve_{ijk}\,H^a_{ij}
\nonumber
\er
Using \rf{conformaltransf}-\rf{fieldtransf5}, and the fact that \rf{conformaltransf} implies that $\partial_j^2\zeta_i=-\partial_iD$,  one gets that 
\be
\delta \Lambda^{(1)}_a=-3\,D\,\Lambda^{(1)}_a\;;\qquad\qquad \qquad\qquad
\delta \Lambda^{(2)}_{ab}=-2\,D\,\Lambda^{(2)}_{ab}
\lab{eqofmotionconformal}
\ee
In addition, one gets that 
\be
\delta \Lambda^{(3)}_{a\,i}=-2\,D\,\Lambda^{(3)}_{a\,i}-\sum_{j\neq i}\partial_i\zeta_j\,\Lambda^{(3)}_{a\,j}
\lab{trickone}
\ee
When calculating \rf{trickone} it is easier to fix the index $i$ to each one of its values, i.e.  $i=1,2,3$. 

Therefore, from \rf{eqofmotionconformal}, one concludes that the static version of the Euler-Lagrange equations \rf{elforu} and \rf{elforh}, are conformally invariant. In addition, from  \rf{trickone}, one observes that the self-dual (BPS) equations  \rf{selfdualeqs} are also conformally invariant.

\section{Proof of relations \rf{uconstantresult}, \rf{eigenvalueshtau} and  \rf{eigenvalueshtaurel}}
\label{app:proof}
\setcounter{equation}{0}

Using \rf{hdef} and \rf{ourstraintensor} one can write the static version of the Euler-Lagrangians equation for $h$ \rf{elforh} as
\br
m_0^2\,e_0^2\,\tau_{ab}=\frac{1}{2}\,h^{-1}_{ac}\,h^{-1}_{bd}\,\ve_{cef}\,\ve_{d{\bar e}{\bar f}}\,\tau_{e{\bar e}}\,\tau_{f{\bar f}}
\lab{elforhtauh}
\er
Suppose now that we diagonalise $h$ with an orthogonal matrix $M$ as
\be
h=M\,h^D\,M^T\;;\qquad\qquad\qquad {\rm with}\qquad \qquad h^D_{ab}=\vp_a\,\delta_{ab}\;;\qquad\qquad M\,M^T=\one
\lab{mdefappendix}
\ee
We are assuming that the eigenvalues $\vp_a$ of the matrix $h$ are all positive. 
Conjugating both sides of the $SU(2)$ commutation relations \rf{su2killing} with a $SU(2)$ group element $g$, and using the definition of the adjoint representation \rf{adjoint}, we get that $\ve_{abc}\,d_{dc}\(g\)=\ve_{efd}\,d_{ea}\(g\)\,d_{fb}\(g\)$. The matrices $d\(g\)$ of the adjoint representation are real and unitary and so they are or\-tho\-go\-nal. Therefore, any orthogonal matrix $M$ satisfy
\be
\ve_{abc}\,M_{dc}=\ve_{efd}\,M_{ea}\,M_{fb}
\lab{epsilonrelform}
\ee
Using \rf{mdefappendix} and \rf{epsilonrelform} in \rf{elforhtauh} one gets
\be
m_0^2\,e_0^2\,\(M^T\,\tau\,M\)_{ab}=\frac{1}{2}\, \frac{1}{\vp_a\,\vp_b}\, \ve_{acd}\,\ve_{bef}\,\(M^T\,\tau\,M\)_{ce}\,\(M^T\,\tau\,M\)_{df}
\lab{elforhtauh1}
\ee
Introduce the real and diagonal matrix $B$ as
\be
B= \mid m_0\,e_0\mid\,{\rm diag.}\,\( \sqrt{ \vp_2\,\vp_3}\,,\, \sqrt{\vp_1\,\vp_3}\,,\, \sqrt{ \vp_1\,\vp_2}\,\)
\lab{cdefappendix}
\ee
and define a symmetric and real matrix $A$ as
\be
B\,A\,B=M^T\,\tau\,M
\lab{cataurel}
\ee
Then the six equations in \rf{elforhtauh1} can be written as
\be
A= A_C
\lab{acofactorrel}
\ee
where $A_C$ is the matrix of cofactors of $A$, which is symmetric since $A$ is symmetric. We have to consider now two distinct cases:
\begin{enumerate}
\item {\bf The case where ${\rm {\bf det}}\,A \neq 0$, in a domain ${\cal D}$ of $\IR^3$}. Then, \rf{acofactorrel} can be written as
\be
A=\({\rm det}\,A\)\; A^{-1}
\ee
Clearly that implies that
\be
{\rm det}\, A=1 \qquad\qquad \mbox{\rm and so}\qquad\qquad A=A^{-1}
\ee
As a consequence of the fact that $A$ equals its inverse, it follows that its eigenvalues are $\pm 1$. But the condition of unity determinant implies that either all eigenvalues are $1$, or one eigenvalue is $1$ and the other two are $-1$. But we have shown in \rf{positivetau} that $\tau$ is a positive matrix, and so is $M^T\,\tau\,M$. Then \rf{cataurel} implies that $B\,A\,B$ is also positive and so is $A$. Consequently, we conclude that all eigenvalues of $A$ are equal to $1$, and so $A=\one$. Then from \rf{cataurel} we have that $M^T\,\tau\,M$ is diagonal. That proves the second relation in \rf{eigenvalueshtau}, once the first is assumed. It then follows that \rf{cdefappendix} and \rf{cataurel} imply that the eigenvalues of $\tau$ are indeed given by \rf{eigenvalueshtaurel}. 

\item {\bf The case where ${\rm {\bf det}}\,A = 0$, in a domain ${\cal D}$ of $\IR^3$}. Then, \rf{cataurel}  implies that ${\rm det}\,\tau={\rm det}\,A\,\({\rm det}\,B\)^2=0$, and so the matrix $\tau$ is singular. But there is more to it. From \rf{acofactorrel} one gets 
\be
A^2= A\,A_C= \({\rm det}\,A\)\;\one=0
\lab{asquarezero}
\ee
But for a symmetric matrix $A$, the diagonal elements of $A^2$ are sums of squares, i.e. $(A^2)_{11}=A_{11}^2+A_{12}^2+A_{13}^2$, and so on. Therefore, as $A$ is real, one concludes from \rf{asquarezero} that $A=0$, and from \rf{cataurel} one gets that $\tau=0$, and so  the whole matrix $\tau$ vanishes identically in ${\cal D}$.  Consequently, from \rf{ourstraintensor} one concludes that $R_i^a=0$ for all values of $a$ and $i$, and so from \rf{rdef} the $SU(2)$ group element $U$ have to be constant in ${\cal D}$. 

\end{enumerate}

\subsection{The analysis using the self-duality equations}
\label{subsec:tausing}

The analysis of the previous subsection has used the static version of the Euler-Lagrange equations \rf{elforh} associated to the $h$-fields, that lead to equations \rf{elforhtauh} and \rf{elforhtauh1}. Such analysis  needed the assumption that the matrix $h$ is invertible and its eigenvalues are strictly positive. However, we can make a similar analysis of the self-dual sector using the self-duality equations \rf{selfdualeqs}.  Contracting \rf{selfdualeqs} with $R_i^c$ and using \rf{ourstraintensor} and \rf{nicedet} we get that
\be
\lambda \,h_{ab}\,\tau_{bc}=\delta_{ac}\, {\rm det}\,R
\ee
Using \rf{signlambdadetr} and the fact that ${\rm det}\,\tau=\({\rm det}\,R\)^2$, we get
\be
\mid \lambda\mid\, h\cdot \tau=\sqrt{{\rm det}\,\tau}\;\one
\lab{htautruerelation}
\ee
We then have some possibilities:
\begin{enumerate}
\item If ${\rm det}\,\tau\neq 0$ in some points or in a region ${\cal D}$ of the physical space, then  we get from \rf{htautruerelation} the relation \rf{goodrelhtau}, and so the matrix $h$ is completely determined from the matrix $\tau$, and consequently from the $SU(2)$ $U$-fields. Therefore, the self-duality equation has a solution in ${\cal D}$, for any configuration of the $U$-fields such that $\tau$ is invertible.  
\item If ${\rm det}\,\tau= 0$ in some points or in a region ${\cal D}$ of the physical space, then \rf{htautruerelation} becomes 
\be
h\cdot \tau=0
\lab{htauzero}
\ee
 Given the $U$-fields and so the $\tau$ matrix, we diagonalise it with an orthogonal transformation $\tau=N\cdot \tau_D\cdot N^T$, with $N\cdot N^T=\one$. Then  \rf{htauzero} becomes 
\be
\(N^T\cdot h\cdot N\)_{ab}\,\omega_b=0\qquad\qquad\qquad \mbox{\rm no sum on $b$}
\lab{htauzero2}
\ee
where $\omega_b$ are the eigenvalues of $\tau$, and $\(N^T\cdot h\cdot N\)$ is a symmetric matrix that has the same eigenvalues as $h$. We now have the following possibilites:
\begin{enumerate}
\item If $\tau$ has just one zero eigenvalue, let us say $\omega_c=0$, then \rf{htauzero2} implies that the rows and columns of $\(N^T\cdot h\cdot N\)$ not corresponding to $c$ must vanish, i.e.
\be
\(N^T\cdot h\cdot N\)_{ab}=0 \qquad\qquad\qquad  b\neq c\qquad {\rm any}\; a
\lab{nullcolumn}
\ee
By the symmetry of $\(N^T\cdot h\cdot N\)$, it follows that \rf{nullcolumn} also holds true  for $a\neq c$ and any $b$.
So, it has  just one non-vanishing element, namely $\(N^T\cdot h\cdot N\)_{cc}$, and that is not determined by the self-duality equation. Therefore, $h$ has to have two zero eigenvalues, and the third one is  undetermined.
\item If $\tau$ has two zero eigenvalues, and so only one non-vanishing eigenvalue, let us say $\omega_c\neq 0$, then \rf{htauzero2} implies that the row and column of $\(N^T\cdot h\cdot N\)$ corresponding to $c$ must vanish, and the remaining $2\times 2$ block is undetermined. Therefore, $h$ has to have one zero eigenvalue and the other two are undetermined. 
\item If $\tau$ has three zero eigenvalues, then $\tau=0$ in ${\cal D}$. In such a case $h$ is completely undetermined. Indeed, from \rf{ourstraintensor} we have that $\tau_{aa}={R_1^a}^2+{R_2^a}^2+{R_3^a}^2=0$, for $a=1,2,3$, and so we conclude that the whole matrix $R_i^a$ vanishes in ${\cal D}$, and the $U$-fields must be constant there. Therefore, the self-duality equations  \rf{selfdualeqs} are satisfied in ${\cal D}$, by any matrix $h$. 
\end{enumerate}
\end{enumerate}

Therefore, we conclude that the more singular the matrix $\tau$ is, the fewer restrictions the self-duality equations  \rf{selfdualeqs} impose on the matrix $h$. Consequently, for any configuration of the $U$-fields (with $\tau$ singular or not)    we always have a matrix $h$ that solves the self-duality equations. For the cases where ${\rm det}\,\tau=0$, the matrix $h$ may not be uniquely determined.

\end{document}